# Hallmarks of the Mott-Metal Crossover in the Hole Doped J=1/2 Mott insulator $Sr_2IrO_4$

Yue Cao[1,*], Qiang Wang[1,#], Justin A. Waugh[1], Theodore J. Reber[1,&], Haoxiang Li[1], Xiaoqing Zhou[1], Stephen Parham[1], Nicholas C. Plumb[2], Eli Rotenberg[3], Aaron Bostwick[3], Jonathan D. Denlinger[3], Tongfei Qi[4], Michael A. Hermele[1], Gang Cao[4], Daniel S. Dessau[1,*]

[1] Department of Physics, University of Colorado, Boulder, CO 80309, USA

[2] Swiss Light Source, Paul Scherrer Institut, CH-5232 Villigen PSI, Switzerland

[3] Advanced Light Source, Lawrence Berkeley National Laboratory, Berkeley, CA 94720, USA

[4] Department of Physics and Astronomy and Center for Advanced Materials, University of Kentucky, Lexington, University of Kentucky, Lexington, KY 40506, USA

# Current address: Los Alamos National Laboratory, Los Alamos, NM 87545, USA

& Current address: Condensed Matter Physics and Materials Science Department, Brookhaven National Laboratory, Upton, NY 11973, USA

* Corresponding authors. Email: Y.C. (ycao@colorado.edu) and D.S.D. (Dessau@colorado.edu)

**The physics of doped Mott insulators [1] remains controversial after decades of active research, hindered by the interplay among possible competing orders and fluctuations [2, 3]. It is thus highly desired to distinguish the intrinsic characters of the Mott-metal crossover from those of other origins. We investigate the evolution of electronic structure and dynamics of the hole-doped J=1/2 Mott insulator $Sr_2IrO_4$. The effective hole doping is achieved by replacing Ir with Rh atoms, with the chemical potential immediately jumping to or near the top of the lower Hubbard band. The doped iridates exhibit multiple exotic features previously observed in doped cuprates [2, 3, 4, 5] – pseudogaps, Fermi "arcs", and marginal-Fermi-liquid-like electronic scattering rates, despite different mechanisms that forbid electron double-occupancy [1, 2, 6, 7]. We argue these universal features of the Mott-metal crossover are not related to preformed electron pairing [8], quantum criticality [9] or density-wave formation [10] as most commonly discussed. Instead, short-range antiferromagnetic correlations may play an indispensible role.**

Finding universal features in interacting electronic systems is a major theme in modern condensed matter research. By universality we refer to the low-energy properties that do not depend on the details of the interactions. For materials with relatively weak electron correlations, the low-energy excitations are well described by the Fermi liquid theory. For doped Mott insulators where correlations are strong, pinpointing the hallmarks common to all Mott-metal crossovers [1, 2, 3] has proven a formidable task. This could largely be attributed to the long candidate list of competing electronic orders, including long-range magnetic order and Fermi surface instabilities, among others, which yield a complex global doping-temperature phase diagram. For example, while the past three decades have witnessed tremendous progress in characterizing the phenomenology of the high-Tc cuprates, the interpretation of most of these experimental observations remains highly controversial. It has become essentially inseparable whether such exotic phenomena as pseudogaps [3, 4] and marginal Fermi liquid scattering rates [5] arise from the metal-insulator transition, certain density-wave instabilities or are fluctuations of the superconductivity. Moreover, the charge insulation in most known Mott insulators arises solely from the Coulomb repulsion. It would be highly desired to study new doped Mott insulators, especially those with a cleaner phase diagram (thus fewer competing orders) and a different mechanism that forbids electron double-occupancy.

$Sr_2IrO_4$ has attracted much interest recently as a new family of Mott insulator [6, 7], with the potential of realizing novel phases of matter [11] and achieving higher superconducting $T_C$ [12]. The Ir-O planes are similar to the Cu-O planes in cuprates, with the Ir atoms sitting at the center of the Ir-O octahedra, except for the 22° Ir-O-Ir bond angle (Figure 1a). There are 5 $t_{2g}$ electrons per Ir. The Ir $t_{2g}$ level splits into the $J_{3/2}$ doublet (filled with 4 electrons) and the $J_{1/2}$ singlet as a result of the strong spin-orbit coupling. The half-filled $J_{1/2}$ further splits into lower (filled with 1 electron) and upper (empty) bands, with this splitting general believed to be due to the Coulomb repulsion, which is why these bands are considered upper and lower Hubbard bands (UHB and LHB, see Figure 1c1) [7]. The insulating behavior in $Sr_2IrO_4$ derives from the coupling of strong spin-orbit interaction with Coulomb repulsions, which is drastically different from that in cuprates. However, the $Sr_2IrO_4$ parent compound, just like undoped cuprates, is antiferromagnetically ordered [13]. Driving $Sr_2IrO_4$ towards metallicity thus provides a unique opportunity to investigate the universal features of the Mott-metal crossover.

Different approaches of doping $Sr_2IrO_4$ and related compounds have been found [14, 15, 16, 17, 18], and of special interest is the $Sr_2Ir_{1-x}Rh_xO_4$ series. With as little as 4% Rh substitution, the normalized resistivity drops by $10^4$. The long-range magnetic order decays more slowly and still survives with a $T_N$~17K for 15% Rh (see Supplementary Information for details). Rh is directly above Ir in the periodic table, so is expected to be isovalent. It has been proposed that the metallicity in the Rh doped iridates comes from the reduced spin-orbit coupling of Rh (due to the smaller atomic number) which then leads to the reduced splitting of the $J_{3/2}$ and $J_{1/2}$ bands [16, 17] as well as the formation of in-gap states [17] (Figure 1c3). As we will show below, the Rh atoms in fact act as effective hole dopants to $Sr_2IrO_4$ (Figure 1c2). So far, no superconductivity has yet been reported in these Rh doped compounds, which is different from doped cuprates. The absence of superconductivity reduces possible competing orders and makes the Rh doped iridates a cleaner system to study – the long-range canted antiferromagnetism (AF) is the only confirmed order in the system.

Angle-resolved photoemission spectroscopy (ARPES) proves an invaluable tool for directly observing both electronic structure and low-energy electron dynamics in doped Mott insulators [3] and bears implication for electronic order including density-wave instabilities. We performed ARPES on single crystals of Rh doped $Sr_2IrO_4$. Details of the sample growth and ARPES experiments are presented in the Methods section and Supplementary Information. Due to the $\sqrt{2} \times \sqrt{2}$ lattice reconstruction, the Brillion zone (BZ) of $Sr_2IrO_4$ and its Rh doped compounds is reduced in half, similar to the formation of the antiferromagnetic BZ in the parent cuprates. We show the constant energy surfaces of $Sr_2IrO_4$ (Figure 1b left panel) and as a comparison, of Pb doped $Bi_2Sr_2CaCu_2O_{8+\delta}$ (Figure 1b right panel). The folded and "original" BZs for the iridates are marked in white solid and blue dashed lines, respectively. The "original" BZ corresponds to one Ir-O plaquette in real space, as shown by the blue dashed line in the left panel in Figure 1a. As we will show later in the paper, the close-to-$E_F$ features of the doped iridates are best captured not by the folded BZ, but instead by the "original" blue BZ. To avoid confusion, we use ($\pi$, 0) to mark the X point in the BZ as defined in [7], and use $\Gamma$' to denote ($\pi$, $\pi$), which is the $\Gamma$ point in the 2nd folded BZ. In Figure 2a-2b, we show the constant energy surfaces as a function of binding energy for $Sr_2IrO_4$ and $Sr_2Ir_{1-x}Rh_xO_4$ with $x$=15% at T=50K. While there is no Fermi surface for the parent compound, there are states at $E_F$ in the $x$=15% iridate, corresponding to enhanced conductivity in the Ir-O plane. The constant energy surface of the $x$=15% compound is quite similar to that of the parent, except that it is shifted in binding energy by ~200meV. To identify the Fermi surface topology, we plot the ARPES spectrum along $\Gamma$'-($\pi$, 0) for both samples (Figure 2c) (along the yellow lines in Figure 2a and 2b). There is a hole-pocket centered at ($\pi$, 0), which comes from the $J_{1/2}$ LHB [7, 19]. The top of the valence band is ~170meV below $E_F$ in the parent compound, and is above $E_F$ for the $x$=15% sample. Indeed, both the $J_{3/2}$ band (white dashed line in Figure 2c) and $J_{1/2}$ LHB (green dashed line) [20, 21] are shifted by ~200meV. It appears that rather than a reduced splitting of the $J_{3/2}$ and $J_{1/2}$ bands, $Sr_2IrO_4$ is hole doped with Rh substitution. Note that while the Rh doped compound displays strong spectral weight extending towards ($\pi$/2, $\pi$/2) (the M point) near $E_F$, the band dispersion at the M point lies below $E_F$ (see Supplementary Information). Therefore, the "Fermi surface" is made up only of the states encircling X or ($\pi$, 0), i.e. it encompasses holes.

We determined the chemical potential shift quantitatively from the valence bands, as shown in Figure 3a. With the increase of Rh concentration, the chemical potential is pushed deeper into the $J_{1/2}$ LHB, confirming that Rh acts as an effective hole dopant. We extrapolate the chemical potential shift at finite Rh densities and derive a ~170meV intercept in the zero doping limit. Note the top of the valence band in the parent compound locates at ($\pi$, 0) and is around 170meV below $E_F$, from both ARPES [19] and scanning-tunneling [22] experiments. Thus upon effective hole doping the chemical potential immediately jumps to the edge of the lower Hubbard band (Figure 1c2), as opposed to competing models (Figure 1c3) [17, 23] where new quasiparticle-like states emerge in the middle of the Hubbard gap. The doping schematic in Figure 1c2 also agrees with the recent optical conductivity measurements [17], while the interpretation in Ref. [17] (Figure 1c3) could not explain the ARPES measured band structure (see Supplementary Information).

The effective hole doping is quite plausible when considering the simple atomic model depicted in Figure 3b. Rh atoms have smaller spin orbit coupling than Ir, leading to the smaller splitting between the $J_{3/2}$ and $J_{1/2}$ states. Assuming the average energy of all 6 $t_{2g}$ states is similar for both Rh and Ir, the empty $J_{1/2}$ state of Rh would then have a lower energy than that of Ir. Thus a $J_{1/2}$ electron from a neighboring Ir atom will fill the $J_{1/2}$ state on the Rh site, leaving behind a hole on the Ir site, as well as a filled and immobile $Rh^{3+}$ site. Of course this is a simplistic model that neither takes into account the finite bandwidth of the effective J states nor the Coulomb repulsion U. Recent X-ray absorption experiments at the Rh $L_3$-edge have confirmed Rh indeed has a valence of 3+ in these compounds [24].

Hereafter we focus on the low-energy electronic dynamics of these hole-doped compounds in search for "universal" features during the Mott-metal crossover. In Figure 4a we show the Fermi surface topology for the $x$=15% sample at 50K. The segments of the Fermi surface centered at $\Gamma$ ($\Gamma'$) and equivalent k locations are highlighted with solid yellow (blue) lines, and labeled FS1 and FS2, respectively. Energy distribution curves (EDCs) from many different k points on the Fermi surface are plotted in Figure 4b. There is a dramatic difference between the EDCs from FS1 and FS2 – those from FS2 are generally pushed away from $E_F$. We use the standard "midpoint of leading edge" method [25, 26] to quantify the spectral weight suppression, by fitting the EDCs to a shifted, broadened leading edge (see Methods section and Supplementary Information), with the amount of shift (defined as the gap size) giving the 50% point of the leading edge. Note that in contrast to the case of a BCS-like gap that works well for superconductivity or a standard charge or spin density wave gap, this shifted edge does not have a pile-up of spectral weight beyond the gap edge, like many other pseudogaps [4, 26], i.e. it does not enforce spectral weight conservation upon the opening of the gap. This absence of quasiparticle peaks in the doped iridates is one of the signatures of their non-Fermi liquid nature. We assign the gap size $\Delta_1$ and $\Delta_2$ from the leading edge fitting to FS1 and FS2 respectively. For $x$=15%, $\Delta_1$ as defined by this leading edge method vanishes, and FS1 can be considered a "regular" piece of Fermi surface (albeit without quasiparticle peaks), while EDCs from FS2 show a partial depletion of near-$E_F$ spectral weight. As FS1 is only topologically connected to the "gapped" FS2, we describe the FS1 as a Fermi surface "arc" and FS2 as "pseudogapped".

We track how $\Delta_1$ and $\Delta_2$ evolve with reduced Rh concentration, as the material gets closer to the Mott insulator. At $x$=4% (Figure 4c), both $\Delta_1$ and $\Delta_2$ are finite, indicating that both FS1 and FS2 are pseudogapped, with $\Delta_1$~3meV and an increased $\Delta_2$~38meV. In Figure 4d we plot the doping dependence of $\Delta_1$ and $\Delta_2$. For $x$=4%~<11%, the entire Fermi surface is pseudogapped, which resembles the ultra-underdoped cuprates [27, 28]. We mark the presence of both pseudogaps as suppressed spectral weight near $E_F$ (the "notch") in Figure 1c2. The pseudogaps observed here are not to be mistaken for the matrix element effect as these gaps happen over a very narrow binding energy range that is essentially identical for different photon energies (see Supplementary Information).

The pseudogap phase in correlated electron systems is often considered a symmetry-broken phase of matter. Thus the origin of the pseudogap could be reflected in its k-space symmetry as well as its thermal evolution. For Rh concentration 4%~15%, $\Delta_1$ ($\Delta_2$) is roughly independent of **k** along the Fermi surface segment FS1 (FS2). In this sense, the pseudogaps in the non-superconducting Rh-doped iridates are clearly different from the near-nodal prepairing pseudogap in the near-optimal cuprates [29], where pseudogaps follow the superconducting paring symmetry.

Figure 4e shows the temperature dependence of EDCs at $k_F$ from FS1 and FS2 for the $x$=11% sample, with the temperature range straddling the AF ordering temperature $T_N$=57K. Within the error bar no obvious changes with temperature are observed, indicating that the pseudogap is not directly related to the long range canted AF order. This is further confirmed in the $x$=15% sample (Figure 4b), where $\Delta_2$ is finite at 50K, above $T_N$=17K. This observation clearly suggests the pseudogap is not tied to the long-range magnetic order, and it is likely the pseudogap phase persists down to the zero-temperature quantum ground state in these hole-doped iridates.

Another commonly considered origin of pseudogaps is the density-wave instabilities in the form of Fermi surface nesting, as has been discussed in the manganites [26] and cuprates [10, 30]. In the case of iridates, it is temping to draw nesting vectors such as the white arrows (Figure 4a) between FS2's with the same gap size $\Delta_2$. However, the same ordering vector Q also connects FS1's, yet with a much smaller gap $\Delta_1$. The Fermi-surface nesting scenario does not explain the preference for a larger gap along FS2 than along FS1. It appears the pseudogap in iridates is inconsistent with many influential explanations for the antinodal pseudogap [26, 30] in manganites and near optimally-doped cuprates though it may be more connected with the more-recently observed nodal pseudogap in heavily underdoped cuprates [27, 28].

The non-Fermi liquid nature of these doped iridates is not only reflected in the absence of quasiparticle peaks along the EDC (Figure 4b and 4c), but also in the single-electron scattering rate. In the Fermi liquid theory, the quasiparticle scattering rate grows linearly with the binding energy (and temperature) squared. For up to 15% Rh substitution, as shown in Figure 5b (with raw data shown in Figure 5a) the scattering rates increase roughly linearly with binding energy – a signature of the Marginal Fermi liquid (MFL) [5]. Moreover, from the resistivity curve in Figure 5c, there is a linear relation between the scattering rate and the sample temperature, as highlighted by the black dashed line. Here we have ignored the upturn of the resistivity at low temperatures that is likely due to a localization effect, as has also been observed in most of the underdoped cuprates [31]. The linear MFL scattering rate is one of the most iconic features of the cuprates and other correlated materials such as the ruthenates [32] and has been attributed to a wide range of ideas including quantum critical fluctuations [5] and the break-up of quasiparticles [33].

Combining the observations presented in Figures 4 and 5, the hole doped $Sr_2IrO_4$ apparently shares striking similarities with the doped cuprates, including the presence of pseudogaps and MFL-like scattering rates, despite detailed differences. Our analysis rules out prepairing and density-wave instabilities as accountable for the pseudogap. Also there is no known quantum critical point in the doped iridates. These are some of the most claimed origins of pseudogaps in cuprates. To identify the origin of these hallmarks in the Mott-metal crossover, we set out to find the common features between iridates and cuprates. While structural distortions appear in both materials, their effect is more likely to be static, and less related to the electron dynamics. We also note that for both materials the parent compound has a long-range AF order that is gradually suppressed with carrier doping. While the pseudogap is not tied to the long-range antiferromagnetism in iridates, the short-range AF correlations that underlie the long-range AF order should still survive regardless of the long-range AF order. Indeed short-range AF correlations may be responsible for many of these hallmarks. For example, short-range AF correlations could indicate a widespread distribution of AF interaction strengths and quantum incoherence, giving rise to the lack of quasiparticle peaks and less well-defined gapping near the Fermi level – a possibility also very relevant for the cuprates and other doped Mott insulators.

We would like to briefly comment on recent debates over whether $Sr_2IrO_4$ is a Mott insulator or a Slater insulator [34, 35]. The main disagreement between the two types of insulators is the role of short-range AF correlation. The Slater insulator is a mean-field concept that ignores the short-range correlation, or at best considers them as fluctuations of the long-range order. It is expected the gap in a Slater insulator will diminish with decreasing long-range magnetic order, trending to zero as the phase transition is approached. Experimentally, there is no clear change in the band structure for both the parent compound [19, 36] and the doped iridates (this work) across the onset of long-range magnetism. This suggests the long-range magnetic order is not necessary for the formation of the gap and $Sr_2IrO_4$ is a Mott insulator. Moreover, the smooth crossover from the Mott insulator $Sr_2IrO_4$ to the hole-doped "bad metal", with the emergence of pseudogaps and marginal Fermi liquid behavior similar to doped cuprates, pinpoints the critical role of short-range correlation in the low energy electron dynamics.

**Methods**

All the ARPES, transport and magnetization data were taken from bulk $Sr_2Ir_{1-x}Rh_xO_4$ samples. Single crystals were grown from off-stoichiometric quantities of $SrCl_2$, $SrCO_3$, $IrO_2$ and $RhO_2$ using self-flux techniques. The ARPES experiments were performed at the PGM-A endstation at the Synchrotron Radiation Center (SRC) of University of Wisconsin-Madison, the Beamline 4.0.3 and 7.0.1 ARPES endstations at the Advanced Light Source, Lawrence Berkeley National Laboratory and the Surface and Interface Science (SIS) beamline of the Swiss Light Source at the Paul Scherrer Institut. The samples were cleaved *in situ* with vacuum better than $5 \times 10^{-11}$ Torr. The band structure and low-energy spectra near the Fermi level were taken with $h\nu$=77eV, 80eV, and 90eV, with an energy resolution ~25meV, which is sharp on the scale of the principle spectral features.

The energy distribution curves (EDCs) in Figure 4 are fit to (with $\omega$ being the energy relative to $E_F$):

$$BG + \frac{A + B\omega}{1 + e^{(\omega + \Delta)/k_B T^*}}$$

which is essentially a Fermi function with variable edge width $k_B T^*$ and with the leading edge midpoint shifted from the chemical potential by the gap value $\Delta$. The BG is the background counts and A, B are fitting coefficients.

**Acknowledgements**

Y.C. and D.S.D. acknowledge J.-X. Dai, K. McElroy, D. Reznik, X.-W. Zhang, A. Zunger, G. Arnold, D. Haskel, J. P. Clancy and Y.-J. Kim for insights and discussions. Y.C. also thanks Y.-D. Chuang, M. Bissen, M. Severson for help setting up the experiment. T.F.Q and G.C. acknowledge support by NSF via grant DMR 1265162. M.H. was supported by the U.S. Department of Energy (DOE), Office of Science, Basic Energy Sciences (BES) under Award # DE-FG02-10ER46686. The ARPES data were collected in part from the Synchrotron Radiation Center, University of Wisconsin-Madison, which was initially supported by the National Science Foundation under Award

No. DMR-0537588, and later primarily funded by the University of Wisconsin-Madison with supplemental support from facility Users and the University of Wisconsin-Milwaukee. The ARPES data were also taken from the Advanced Light Source and the Swiss Light Source. The former is supported by the Director, Office of Science, Office of Basic Energy Sciences, of the U.S. Department of Energy under Contract No. DE-AC02-05CH11231.

**Figure Captions**

**Figure 1**

**$Sr_2IrO_4$ as a Mott insulator on the square lattice. a**. The real-space unit cells of $Sr_2IrO_4$ and $La_2CuO_4$. **b**. The k-space unit cells of the same, with matching color scaling and with near-$E_F$ ARPES spectral weight. Ignoring the 22° Ir-O twists gives the blue cells in real and k-space, and corresponds to the regular unit cell of $La_2CuO_4$. Including these twists in $Sr_2IrO_4$ (black, panel **a**) back-folds the k-space cell in k-space (white), similar to the AF order in the parent cuprates. **c1**. The formation of the Mott gap in $Sr_2IrO_4$ as a result of the spin-orbit coupling and Coulomb interaction. **c2**. and **c3**. Possible schematics of chemical potential evolution with Rh doping.

**Figure 2**

**Constant energy surfaces and high symmetry cuts of the parent and Rh doped $Sr_2IrO_4$. a**. and **b**. The constant energy surfaces of the parent (**a**) and $x$=15% Rh substituted (**b**) $Sr_2IrO_4$. The numbers are binding energies relative to $E_F$. The solid white / dashed blue lines are the folded / original BZs. **c**. ARPES energy-momentum intensity plots along $\Gamma'$-$(\pi, 0)$-$\Gamma'$ (yellow lines in panels **a** and **b**). The dashed green and white lines through the data guide the eye for the $J_{1/2}$ and $J_{3/2}$ bands respectively.

**Figure 3**

**Rh atoms act as hole dopants in $Sr_2IrO_4$. a**. Chemical potential shift vs. Rh concentration. The chemical potential shift is measured from the shift of both $J_{1/2}$ LHB and $J_{3/2}$ bands. **b**. Simple atomic picture of hole doping, ignoring band effects and Mott splitting. With a roughly similar average energy for both Ir and Rh sites, the smaller on-site spin-orbit splitting on the Rh sites lowers the $J_{1/2}$ energy relative to that of the host Ir sites. This causes an electron transfer to Rh, i.e. hole doping of the Ir lattice.

**Figure 4**

**Fermi surface segments and pseudogaps in hole doped $Sr_2IrO_4$.** All the data were taken at 50K unless otherwise noted. **a**. Fermi surface spectral weight of the $x$=15% sample, with a hole-like Fermi pocket centered around the ($\pi$, 0) point of the unfolded (blue dashed) BZ. The FS pocket is separated into segments FS1 (yellow) and FS2 (blue), with FS1 facing $\Gamma$ and FS2 facing $\Gamma$'. Q vectors (white arrows) are possible density-wave nesting vectors. **b**. and **c**. EDCs from multiple locations along the FS1 and FS2 segments (yellow and blue, respectively) taken from the $x$=15% and $x$=4% samples. Locations of the individual EDCs are marked by the open colored circles in panel **a**. The leading edges of most EDCs do not reach $E_F$, i.e. they are gapped or pseudogapped. Gap sizes extracted using the "midpoint of leading edge method", are shown in panels **b** and **c** and compiled in panel **d**, with $\Delta_1$ labeling the gaps from FS1 and $\Delta_2$ the gaps from FS2. **e**. EDCs from FS1 (dashed) and FS2 (solid) showing minimal temperature dependence across the magnetic phase transition of the $x$=11% sample.

**Figure 5**

**Marginal-Fermi-liquid-like single-electron scattering rates. a**. The energy-momentum intensity plot along $\Gamma$-X=($\pi$, 0)-$\Gamma$ for an $x$=11% sample at T=50K, passing through four pieces of FS1 Fermi surface. The peak centroids obtained from double Lorentzian fittings to the momentum distribution curves (MDCs) are marked with blue lines. The inset shows where the cut in the main figure is taken from the BZ. **b**. The full-width-half-maximum (FWHM) from the double Lorentzian fitting is plotted vs. the binding energy, showing a linear "marginal Fermi liquid" scattering rate. **c**. Resistivity vs. temperature for the same sample showing a linear dependence at intermediate temperatures. The black dashed line is a guide to the eye.

# Figure 1

**a**

Sr$_2$IrO$_4$   La$_2$CuO$_4$

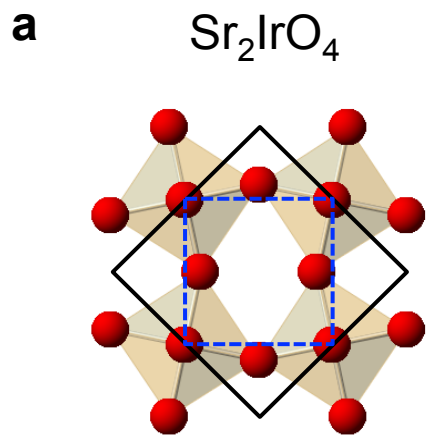 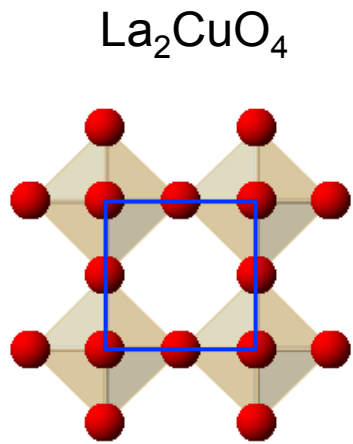

**b**

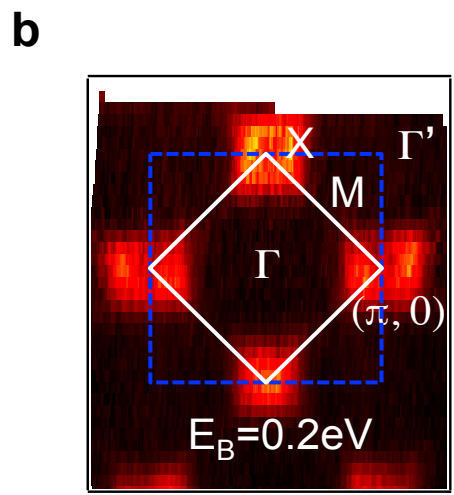 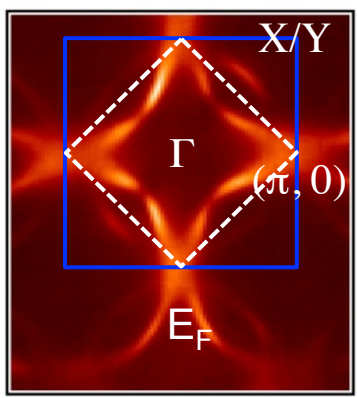

**c1**

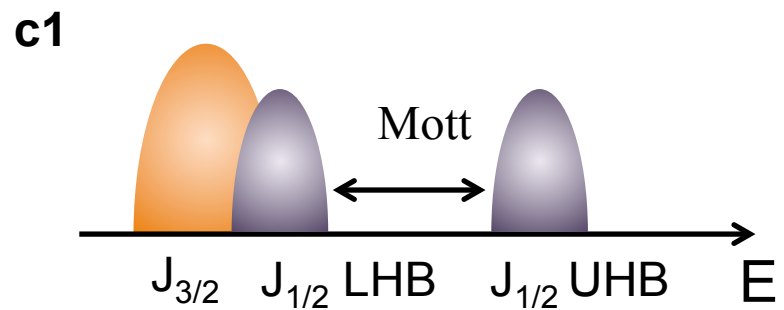

**c2**

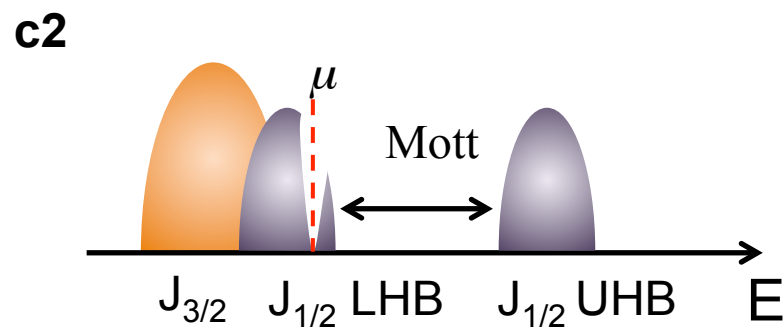

**c3** (inconsistent with ARPES)

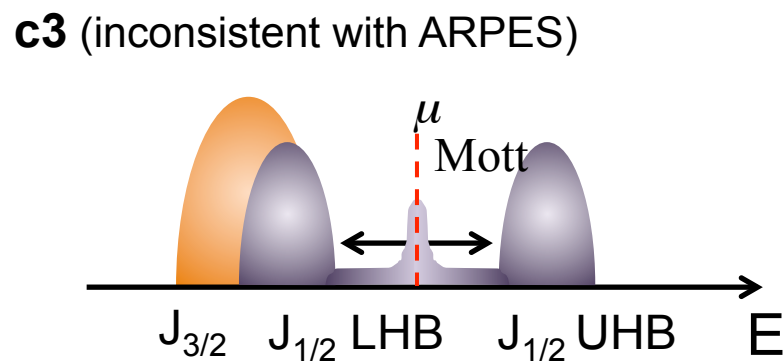

# Figure 2

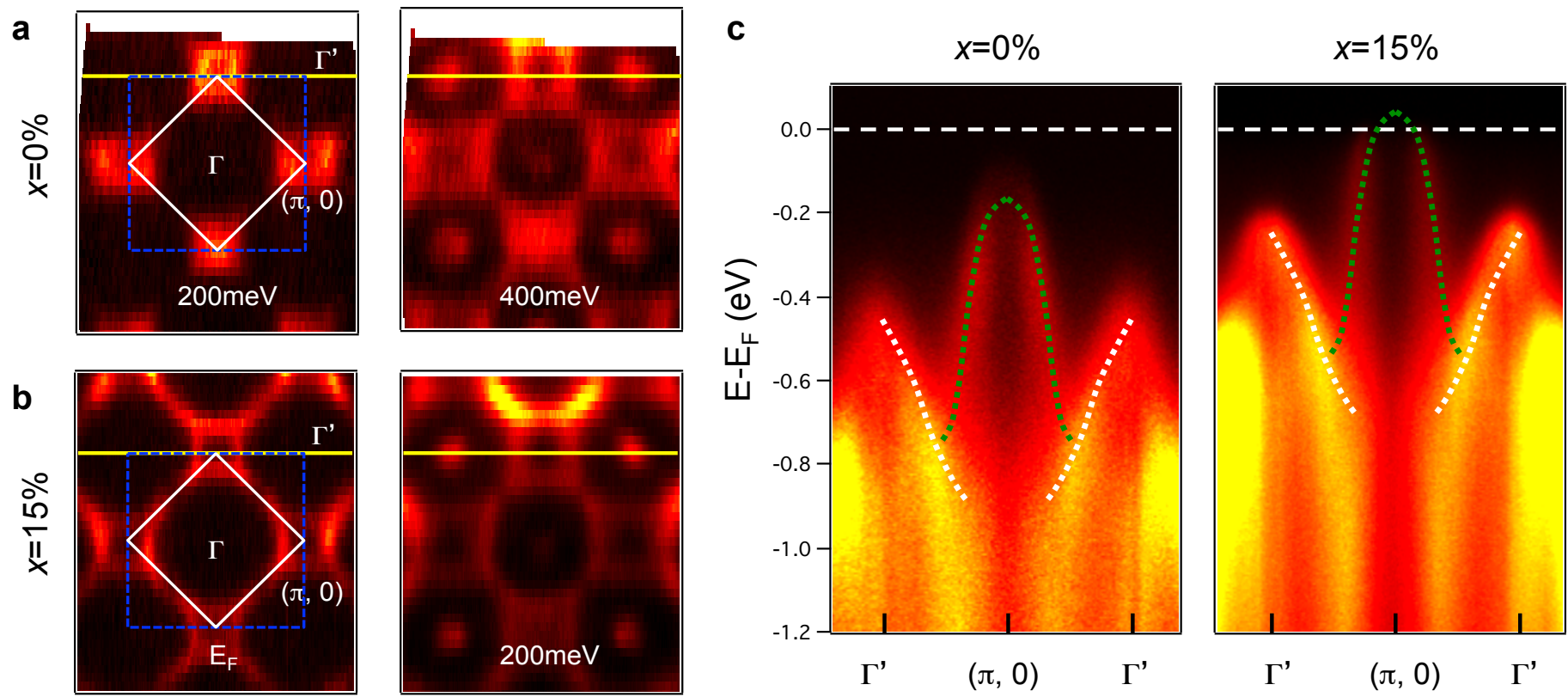

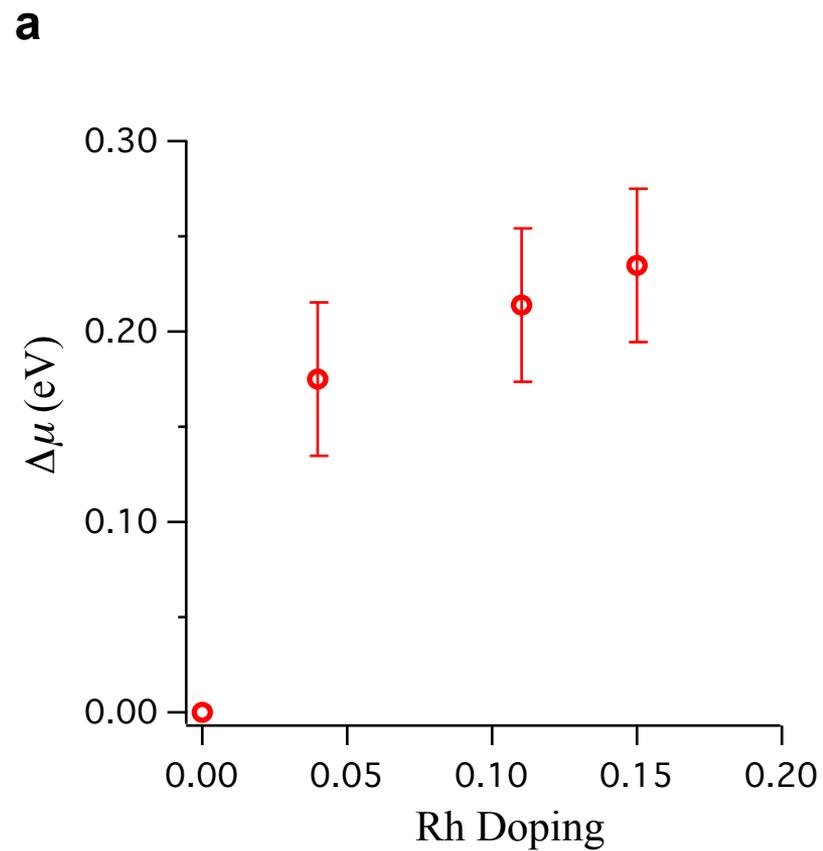
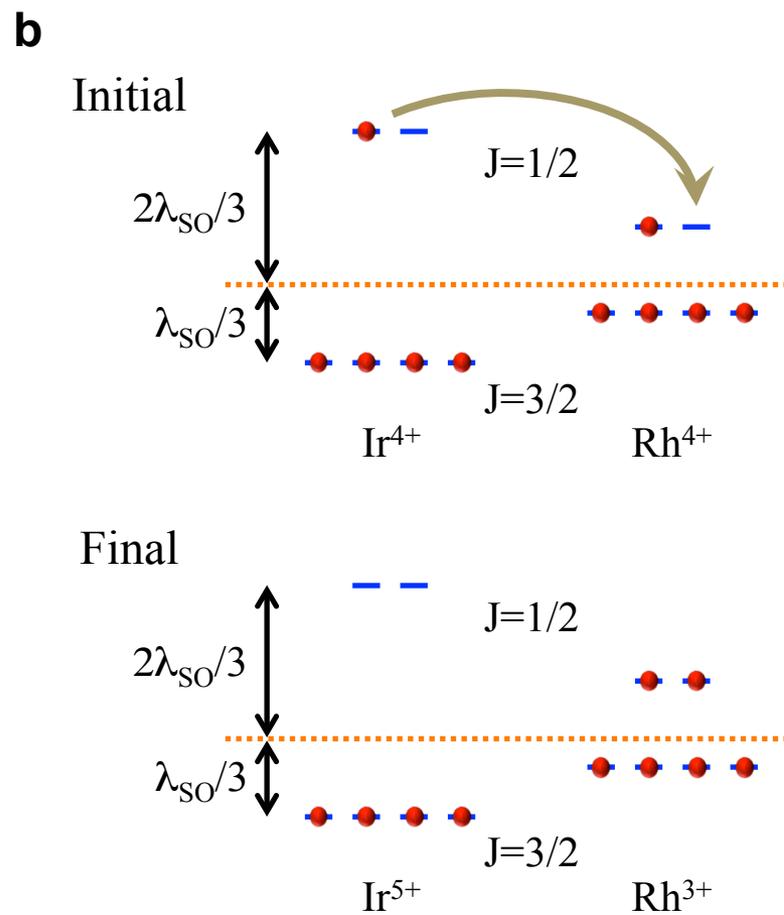

**Figure 4**

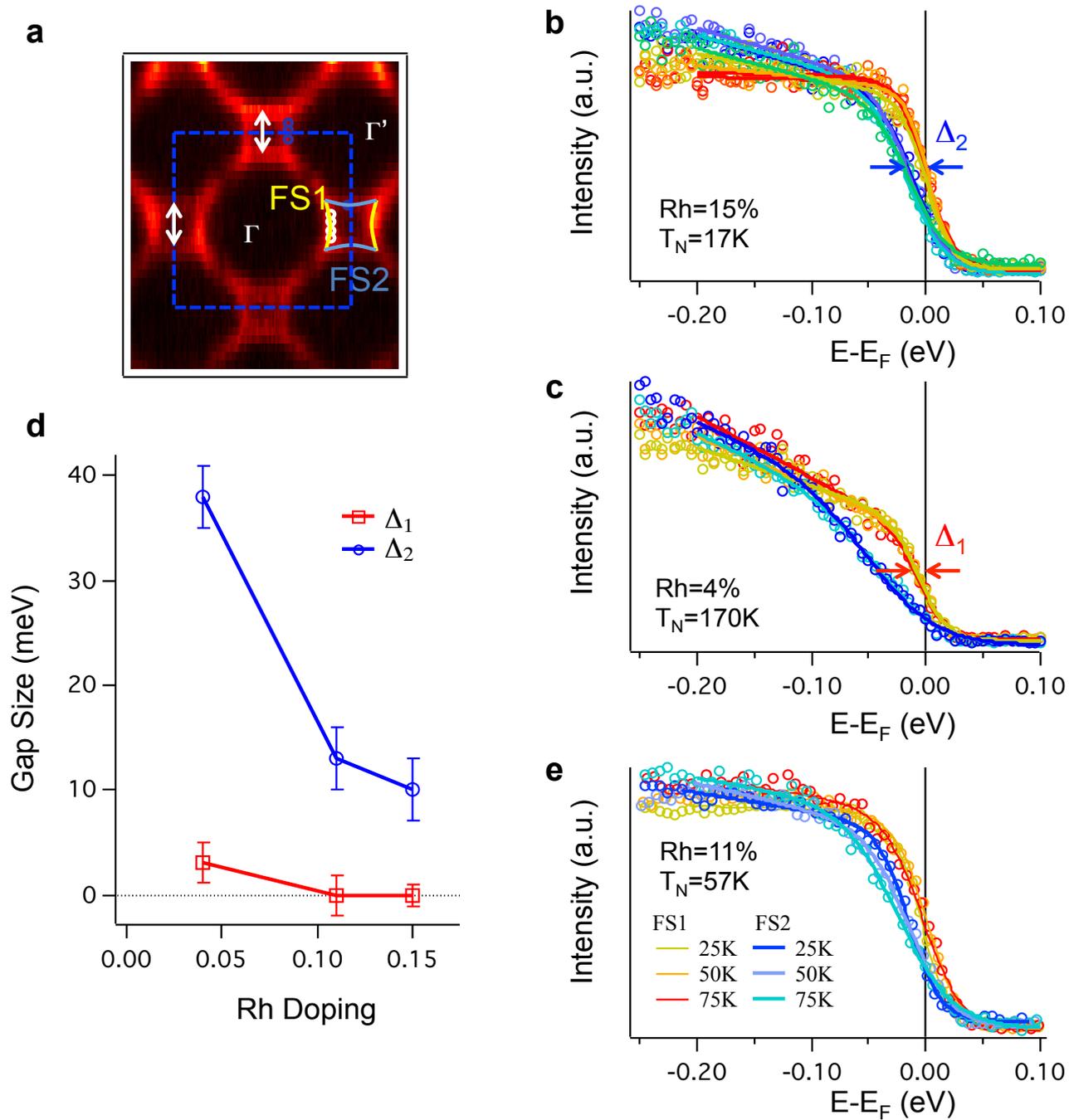

**Figure 5**

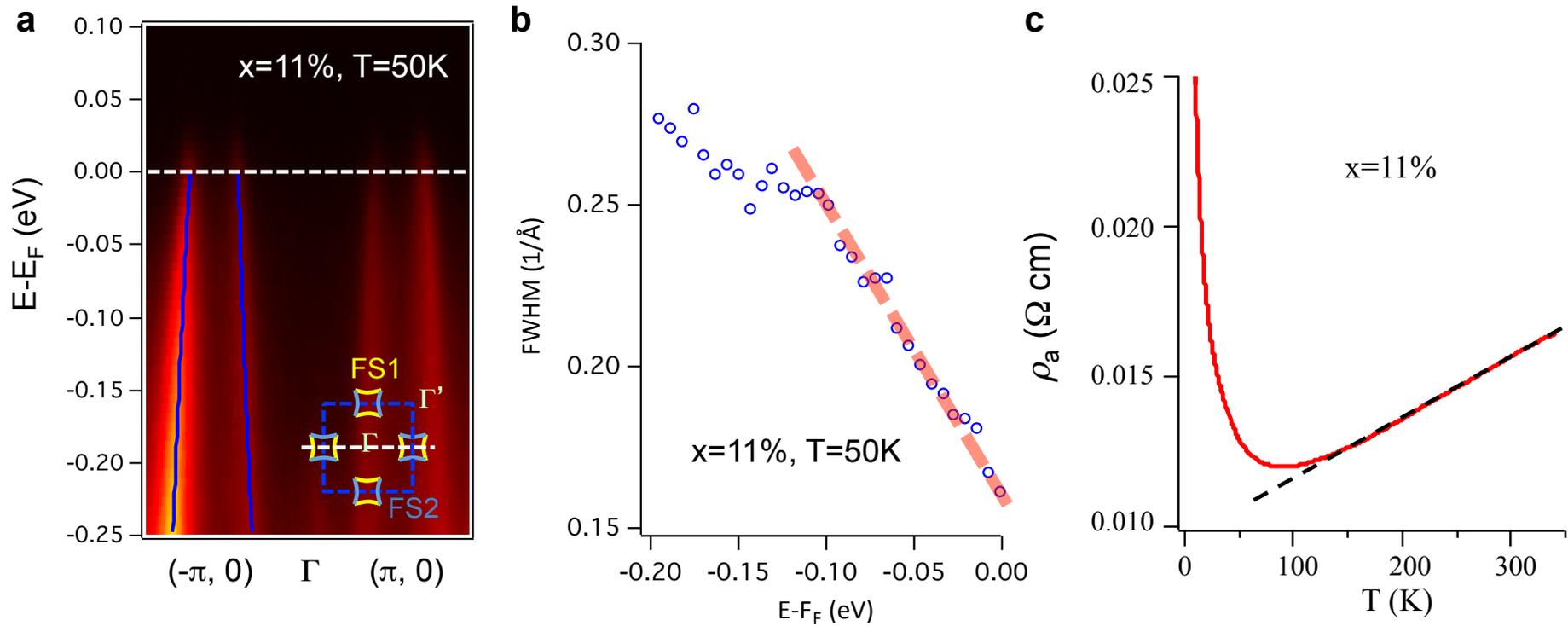

# Hallmarks of the Mott-Metal Crossover in the Hole Doped J=1/2 Mott Insulator $Sr_2IrO_4$
## Supplementary Information

Yue Cao[1,*], Qiang Wang[1,#], Justin A. Waugh[1], Theodore J. Reber[1,&], Haoxiang Li[1], Xiaoqing Zhou[1], Stephen Parham[1], Nicholas C. Plumb[2], Eli Rotenberg[3], Aaron Bostwick[3], Jonathan D. Denlinger[3], Tongfei Qi[4], Michael A. Hermele[1], Gang Cao[4], Daniel S. Dessau[1,*]

[1] Department of Physics, University of Colorado, Boulder, CO 80309, USA
[2] Swiss Light Source, Paul Scherrer Institut, CH-5232 Villigen PSI, Switzerland
[3] Advanced Light Source, Lawrence Berkeley National Laboratory, Berkeley, CA 94720, USA
[4] Department of Physics and Astronomy and Center for Advanced Materials, University of Kentucky, Lexington, University of Kentucky, Lexington, KY 40506, USA

# Current address: Los Alamos National Laboratory, Los Alamos, NM 87545, USA
& Current address: Condensed Matter Physics and Materials Science Department, Brookhaven National Laboratory, Upton, NY 11973, USA
* Corresponding authors. Email: Y.C. (ycao@colorado.edu) and D.S.D. (Dessau@colorado.edu)

## 1. Bulk crystal growth and characterization

All the transport, magnetization and photoemission experiments were performed on bulk $Sr_2Ir_{1-x}Rh_xO_4$ samples. Single crystals studied were grown from off-stoichiometric quantities of $SrCl_2$, $SrCO_3$, $IrO_2$ and $RhO_2$ using self-flux techniques. These mixtures were heated to 1480°C in Pt crucibles, fired for 20 hours, cooled at 5°C/hour to 1440°C, and then rapidly quenched to room temperature. The quenching process is critical to ensure an exclusion of precursor phases such as $Sr_2IrO_4$.

In Figure S1, we show the normalized resistivity (left axis) and the onset temperature of the long-range magnetic order (right axis) as a function of Rh concentration. The long-range antiferromagnetism manifests itself as a weak ferromagnetism in the magnetic susceptibility, due to the canted Ir moment in the Ir-O plane. This is also why we call the onset temperature as the Neel temperature.

It is interesting to note that the insulating behavior melts rapidly (as the resistivity is drawn using log scale), while the long-range magnetic order did not vanish till 15%. The resistivity/magnetism phase diagram in the Rh doped $Sr_2IrO_4$ shares intriguing similarity to the phase diagram, e.g. of $(La_{1-x}Sr_x)_2CuO_4$ [1]. It would seem plausible this kind of phase diagram might be intrinsic in the doped Mott insulators. Further information, including the transport and magnetization measurements are described in Ref. [2].

## 2. ARPES Experiment Setup

The ARPES (angle-resolved photoemission spectroscopy) experiments were performed at the PGM-A endstation at the Synchrotron Radiation Center (SRC) of University of Wisconsin-Madison, the Beamline 4.0.3 and 7.0.1 ARPES endstations at the Advanced Light Source, Lawrence Berkeley National Laboratory and the Surface and Interface Science (SIS) beamline of the Swiss Light Source at the Paul Scherrer Institut. The samples were cleaved *in situ* with vacuum better than $5 \times 10^{-11}$ Torr. The band structure and low-energy spectra near the Fermi level were taken with $h\nu$=77eV, 80eV, and 90eV, with an energy resolution ~25meV.

## 3. Band structure of $Sr_2Ir_{1-x}Rh_xO_4$

We compare the band structures of the Mott insulator $Sr_2IrO_4$ and of the $x$=15% Rh-substituted sample. Due to the local rotation of the Ir-O octahedra, there is a $\sqrt{2} \times \sqrt{2}$ lattice reconstruction in the whole temperature range where ARPES experiments are performed (and in fact up to the room temperature). The Brillouin zone (BZ) is folded for both compounds. To compare to existing first-principles and dynamical mean-field calculations, $(\pi, 0)$ point corresponds to the X point and $(\pi/2, \pi/2)$ the M point. As shown in the main text, while the folded BZ suffices to describe the band dispersion, *only* the "original" BZ corresponding to one Ir-O plaquette captures the low-energy electron dynamics. For this reason we define $(\pi, \pi)$ as the Γ' point, to distinguish it from the Γ point at $(0, 0)$ or $(2\pi, 0)$.

To find the band structure, we took the second derivative of the energy distribution curves (EDCs) along high symmetry directions of the BZ, as shown in Figure S2. This technique is routinely used to highlight the band dispersion in ARPES experiments [3, 4, 5]. We obtain the band structures from the positive peaks on these EDC second-derivative maps, and by comparing band dispersions in the folded and "original" BZs. For example, for cuts along X-Γ-X and Γ'-X-Γ', while they are not equivalent in terms of the low-energy dynamics, e.g. pseudogaps, they have the same large-energy-scale band dispersion. For both the parent compound and the *x*=15% sample, the $J_{1/2}$ lower Hubbard band (LHB) and $J_{3/2}$ band are drawn in dashed green and red lines respectively. The J character of the valence bands is assigned as in [6, 7]. It is evident that both the $J_{1/2}$ LHB and $J_{3/2}$ bands moved towards $E_F$, with neither major relative shift between these two bands nor bandwidth renormalization.

A conservative estimate of the uncertainty of this procedure in determining the peak positions is the half-width of the positive peaks of the second-derivative curves. The chemical potential shift is defined as the average shift in energy by comparing the band structures of the parent compound and the Rh substituted samples. The uncertainty of the chemical potential shift is then the sum of the uncertainties of the peak positions of both the pristine and the doped compounds. For Figure 3a in the main text we have determined these uncertainties at a few high symmetry k points and averaged their results.

## 4.  Comparison to optical conductivity measurements

In Figure 1c2 and 1c3 of the main text we present two possible scenarios how Rh substitution bring metallicity into the parent compound. Figure 1c3 is similar to the proposal in Ref. [8]. In Figure 1c2, the chemical potential "jumps" immediately to the lower Hubbard band with hole doping, while in Figure 1c3 there is in-gap spectral weight transferred from the upper and lower Hubbard bands and the chemical potential is pinned to this in-gap spectral weight.

The optical conductivity experiments could not distinguish between these two scenarios. The $\alpha$ and $\beta$ transition peaks seen in the optical conductivity [8] are interpreted as the optical transition from the $J_{1/2}$ lower Hubbard band and $J_{3/2}$ band to the $J_{1/2}$ upper Hubbard band, respectively. For the Rh concentrations less than 20%, the $\alpha$ and $\beta$ peak locations do not change – an indication there is not much relative shift between the $J_{3/2}$ band and $J_{1/2}$ Hubbard bands. This is consistent with both scenarios in Figure 1c2 and Figure 1c3. The ARPES experiment as presented here, however, unambiguously shows that there is a sudden jump in the chemical potential, thus ruling out the scenario with the in-gap state as shown in Figure 1c3 of the main text. Moreover, the low energy peak that appears below the $\alpha$ peak in the Rh substituted samples could also be explained in Figure 1c2. Since there is suppression of spectral weight near $E_F$ for the Rh doped samples, the low energy peak could originate from the intra-band optical transition across the pseudogap, without introducing in-gap states near the middle of the gap as in Figure 1c3.

## 5. Spectral weight suppression separated from possible matrix element effects

In this section we justify that the low-energy spectral weight suppression (pseudogap) in the Rh doped $Sr_2IrO_4$ is not the result of the "matrix element effect". By "matrix element" [9], we refer to the electron-photon cross-section $I \propto |\langle \psi_f | \vec{A} \cdot \vec{p} | \psi_i \rangle|^2 \propto |\langle \psi_f | \vec{E} \cdot \vec{r} | \psi_i \rangle|^2$. Here $I$ is the measured ARPES intensity, $\vec{A}$ is the electromagnetic gauge and $\vec{p}$ is the electron momentum. $|\psi_i\rangle$ and $|\psi_f\rangle$ are the wavefunctions of the initial state of the electron in the solid and the final state of the photo-excited electron, respectively. For certain combinations of the initial state, final state, and photon dipole operator it is in general possible that the matrix element becomes very small or vanishes, giving small ARPES spectral weight. It is important to be able to confirm that the pseudogaps observed in the iridates are not due to such matrix element effects.

We have measured the doped iridates in up to 8 inequivalent sectors of FS1 and FS2, have used two different sample rotations (($\pi$,0) parallel to the scattering plane and ($\pi$,$\pi$) parallel to the scattering plane), and multiple photon energies (77eV, 80eV and 90eV). In all cases we have observed the same qualitative behavior of a large pseudogap along FS2 and a weak or small pseudogap along FS1. While strongly varying matrix elements might have affected the data at one geometry or photon energy, it is not plausible that they could have affected the data for all experimental conditions we have explored.

Moreover, the spectral weight is suppressed within <50meV of $E_F$ across the whole Rh doping range for both FS1 and FS2. Spectral weight suppression originating from the initial-state orbital selectivity usually appears over a much larger energy scale, which is not consistent with the observed pseudogap size in the hole-doped iridates.

## 6. Quantifying the pseudogap

As discussed in the main text, the EDCs do not have sharp quasiparticle peaks near the Fermi level. Similar broad EDCs have also been observed in lightly doped cuprates [10] as well as manganites [11]. This indicates the effectively hole-doped iridates are certainly not Fermi liquids. To characterize the gap, we use the "midpoint of leading edge" method and fit the EDCs to (with $\omega$ being the energy relative to $E_F$):

$$BG + \frac{A + B\omega}{1 + e^{(\omega+\Delta)/k_B T^*}}$$

which is essentially a Fermi function with variable edge width $k_B T^*$ and with the leading edge midpoint shifted from the chemical potential by the pseudogap value $\Delta$. We use a constant BG to describe the spectral weight above $E_F$, and A, B are fitting coefficients.

For each Rh doping, the pseudogap sizes $\Delta_1$ and $\Delta_2$ are taken to be the average value of more than 10 EDCs on the Fermi surface, along FS1 and FS2 respectively. Accordingly the uncertainty of $\Delta_1$ and $\Delta_2$ are defined as the maximum deviation of the fitted gap size of individual EDCs from the averaged value.

**Figs. S1 and S2**
**Figure Captions**
**Figure S1. The evolution of resistivity and magnetism with Rh doping.** The left axis: the normalized resistivity; and the right axis the onset of the long-range magnetic order.

**Figure S2. The band structure along high symmetry directions in the parent and Rh doped $Sr_2IrO_4$. a.** The Fermi surface topology of the $x$=15% Rh-doped sample. The high symmetry cuts in panels **b** to **d** are drawn in dashed orange line. **b-d.** The EDC second-derivative maps for the parent (**b1-d1**) and $x$=15% Rh-doped sample (**b2-d2**)

respectively. The dashed lines are guides to the eye where the green and red colors denote the J=1/2 lower Hubbard band and J=3/2 band.

# Figure S1

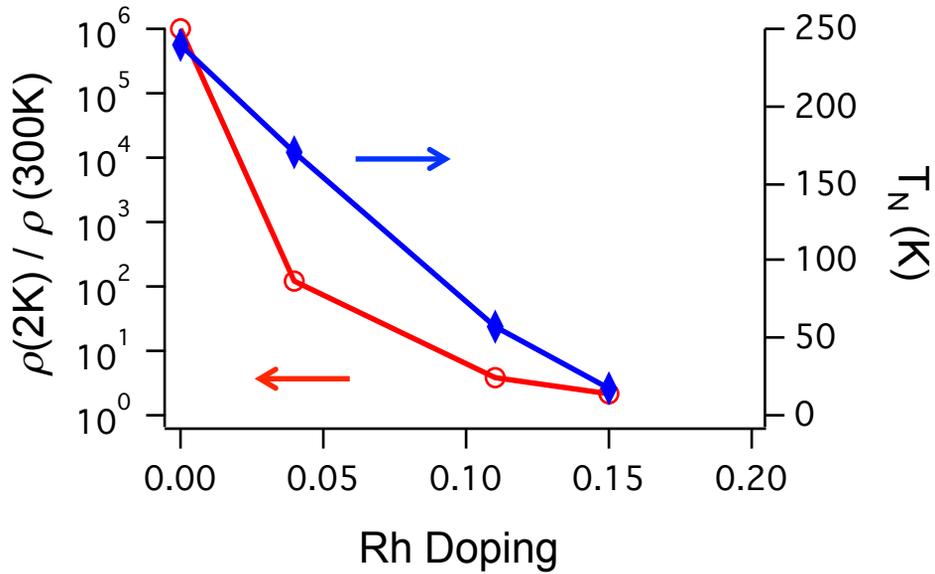

**Figure S1. The evolution of resistivity and magnetism with Rh doping.** The left axis: the normalized resistivity; and the right axis the onset of the long-range magnetic order.

# Figure S2

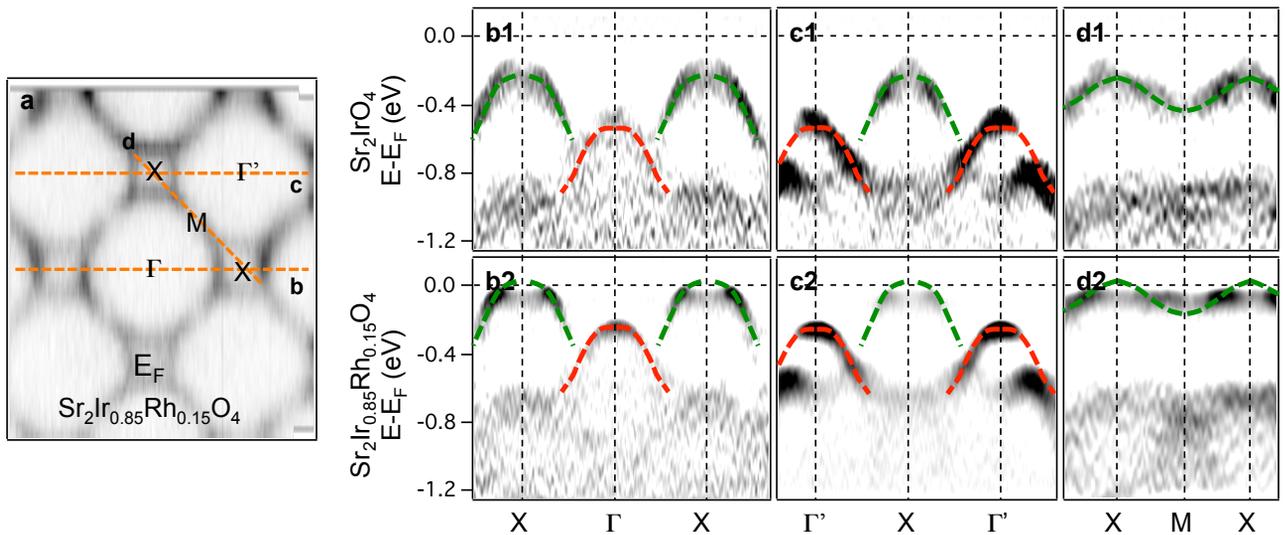

**Figure S2. The band structure along high symmetry directions in the parent and Rh doped $Sr_2IrO_4$. a.** The Fermi surface topology of the x=15% Rh-doped sample. The high symmetry cuts in panels **b** to **d** are drawn in dashed orange line. **b-d.** The EDC second-derivative maps for the parent (**b1-d1**) and x=15% Rh-doped sample (**b2-d2**) respectively. The dashed lines are guides to the eye where the green and red colors denote the J=1/2 lower Hubbard band and J=3/2 band.